\documentclass[aps,prl,preprint,groupedaddress]{revtex4-1}

\usepackage{amsmath,amssymb} \DeclareMathOperator{\sign}{sign}
 
\def\al{\alpha}

\def\ga{\gamma}
\def\de{\delta}
\def\ep{\epsilon}

\def\la{\lambda}

\def\Si{\Sigma}

\def\mn{{\mu\nu}}

\def\frac#1#2{{\textstyle{{#1}\over {#2}}}}

\def\lsim{\mathrel{\rlap{\lower4pt\hbox{\hskip1pt$\sim$}}
    \raise1pt\hbox{$<$}}}
\def\gsim{\mathrel{\rlap{\lower4pt\hbox{\hskip1pt$\sim$}}
    \raise1pt\hbox{$>$}}}
\def\sqr#1#2{{\vcenter{\vbox{\hrule height.#2pt
         \hbox{\vrule width.#2pt height#1pt \kern#1pt
         \vrule width.#2pt}
         \hrule height.#2pt}}}}

\newcommand{\beq}{\begin{equation}}
\newcommand{\eeq}{\end{equation}}
\newcommand{\bea}{\begin{eqnarray}}
\newcommand{\eea}{\end{eqnarray}}

\renewenvironment{thebibliography}[1]
 { \rm
   \begin{list}{\arabic{enumi}.}
    {\usecounter{enumi} \setlength{\parsep}{0pt}
     \setlength{\itemsep}{3pt} \settowidth{\labelwidth}{#1.}
     \sloppy
    }}{\end{list}}
 
\begin{document}
\titlepage

\vglue 1cm
	    
\begin{center}
{{\bf Singular Lorentz-Violating Lagrangians and Associated Finsler Structures.\\}
\vglue 1.0cm
{Don Colladay and Patrick McDonald \\} 
\bigskip
{\it New College of Florida\\}
\medskip
{\it Sarasota, FL, 34243, U.S.A.\\}
 
\vglue 0.8cm
}
\vglue 0.3cm
 
\end{center}
 
{\rightskip=3pc\leftskip=3pc\noindent
Several lagrangians associated to classical limits of lorenz-violating fermions in the 
Standard Model extension (SME) have been shown to yield Finsler functions when the
theory is expressed in Euclidean space. 
When spin-couplings are present, the lagrangian can develop singularities that obstruct the construction of a globally defined Legendre transformation, leading to singular Finsler spaces.   
A specific sector of the SME where
such problems arise is studied.  It is found that the singular behavior can be eliminated by
an appropriate lifting of the problem to an associated algebraic variety.   This provides a smooth classical model for the singular problem. 
In Euclidean space, the procedure involves combining two related singular Finsler functions
into a single smooth function with a semi-positive definite quadratic form defined on a
desingularized variety.} 

\vskip 1 cm

\newpage
 
\baselineskip=20pt
 
{\bf \noindent I. INTRODUCTION}
\vglue 0.4cm

The potential for breaking of Lorentz symmetry in physics underlying the standard model 
has been proposed in a variety of contexts, including a promising mechanism for spontaneous
violation arising within string field theory \cite{kps}.
The Standard Model Extension (SME) involves general Lorentz-violating parameters that can 
arise from such an underlying theory \cite{ck}.
The theory is formulated in the framework of relativistic quantum field theory which involves 
a natural expression in momentum space that leads to dispersion relations that can modify
particle propagation in the classical limit.
The usual procedure for obtaining a classical limit involves performing a Foldy-Wouthuysen
transformation on the underlying Hamiltonian and identifying the new coordinate operator
in the resulting representation as the relevant classical position operator.
At this point, the theory is formulated in terms of momentum, whereas the classical trajectories
are measured in terms of the rate of change of the expectation value of the new position operator,
referred to as the classical particle velocity.
To determine the classical trajectories, it is therefore desirable to find a classical effective 
lagrangian for the theory.
A good analogy is the ray optics limit of Maxwell equations, a formulation of tremendous
value and simplification when the wave nature of light is largely irrelevant.
Such a project was initiated in \cite{kr} with the successful implementation of the an exact 
Legendre transformation for some special choices of Lorentz-violation parameters in the
fermion sector of the SME.
Since then, several other papers have presented various other exactly-solvable cases
and limits \cite{otherlags}.
It was noticed \cite{kosfins} that when converted to Euclidean space, these lagrangians 
generated a variety of Finsler functions, some of which were singular.
It was also pointed out that Finsler space as a generalization to Riemann space may be 
a way to evade the "no go" theorem of including explicit symmetry breaking into general
relativity theories \cite{alangrav}.

In simple cases where spin couplings are irrelevant, either modified Minkowski or Randers
spaces are found to emerge.
On the other hand, spin-dependent couplings produce multiple-valued lagrangians that lead
to a set of singular Finsler functions.
Singular sets are generically present in these functions where the resulting metrics can diverge
and impede the construction of a global Finsler geometry.
In the original paper \cite{kosfins}, these singular sets were simply removed from the space 
leaving a singular Finsler space \cite{shen1} defined on the remaining open subsets.
There are several undesirable features of this approach, the most obvious one being that the resulting space
is not complete, so the particles described by the corresponding lagrangian are forbidden
to travel with certain velocities.  
In some cases, these velocities are not actually attainable and are irrelevant physically, but in
other cases they can be easily accessible creating a serious impediment to formulating the
full theory in terms of Finsler geometry.

A plot of the indicatrix of each singular Finsler function generally reveals cusps at the 
singular sets.
When the singular Finsler functions associated with a specific Lorentz-violating parameter are expressed in terms of a single algebraic variety, the cusps
on the indicatrix are replaced by singularities on the variety.
In this work, we apply a desingularization procedure to resolve the singular points.
The "Finsler b-space" resulting from one of the simplest spin-dependent couplings
is analyzed in detail. 
Note that a recent paper has constructed some interesting classical physics models that lead
to these Finsler functions \cite{ralph} providing intuition about some properties of the space.

\vglue 0.6cm
{\bf \noindent II. FINSLER b-SPACE}
\vglue 0.4cm

One of the first spin-dependent terms in the SME to yield a relatively simple lagrangian was the 
term $ b^\mu \overline \psi \ga^5 \ga_\mu \psi$.
The classical lagrangian corresponding to this field-theoretic term is calculated by performing a Legendre Transformation of the dispersion relation \cite{kr}, with result
\beq
{\cal{L_\pm}} = - m \sqrt{u^2} \mp \sqrt{(b \cdot u)^2 - b^2 u^2} ,
\eeq
where the invariant product was taken to be flat Minkowskian and the $b^\mu$ a constant 
vector field.
In a subsequent work \cite{kosfins}, the theory was extended to a more general setting
where the invariant product is determined by a pseudo-Riemannian metric $r_\mn (x)$ and the one-form $b^\mu (x)$ was extended to a general function of $x$.
This procedure is most likely to work if the fields are slowly varying over spacetime so that effects of
derivatives of the Lorentz-violating background fields and the metric can be neglected in the 
Foldy-Wouthuysen transformation that leads to the classical limit.
For simplicity, we restrict the presentation here to constant background fields with Minkowski
product.
The lagrangian can be Wick rotated to Euclidean space yielding the 
Finsler b-space functions \cite{kosfins}
\beq
F_\pm=\sqrt{y^2}\pm \sqrt{b^2 y^2 - (b \cdot y)^2},
\label{origfins}
\eeq
where $y^i$ are the velocity components in $4$ dimensions (which can easily be generalized to $n$), 
the mass has been set to unity, and the inner product is Euclidean.
It is convenient to separate $y$ into components $y_p$ along $b$, and $y_\perp^i$
 perpendicular to $b$.  
 The quantity under the second square root sign then
reduces to $b^2 y_\perp^2$. 
If one chooses either $F_+$ or $F_-$ to compute the Finsler metrics $g^+$ and $g^-$ as
\beq
g_{ij}^\pm = {1 \over 2} {\partial^2 F_\pm^2 \over \partial y^i \partial y^j} ,
\eeq
the resulting metric components in this special coordinate system are
\beq
g_{00}^\pm = 1 \pm \sqrt{b^2 y_\perp^6 \over y^6} \quad , \quad
g_{0i}^\pm = \pm \sqrt{b^2 y_p^6 \over y^6} ~ \hat y_\perp^i \quad ,
\eeq
\beq
g_{ij}^\pm = \left( 1 + b^2 \pm \sqrt{b^2 \over y^2 y_\perp^2} ~ (y^2 + y_\perp^2) \right) \de^{ij}
\mp \sqrt{b^2 \over y^6 y_\perp^2}y_p^4 \hat y_\perp^i \hat y_\perp^j
\label{metric1}
\eeq
where $0$ indicates the index along $b$ and $i = \{ 1,2,\cdots , n-1 \}$ are the directions 
perpendicular to $b$.

The components in Eq.(\ref{metric1}) have singular behavior along the line $y_\perp = 0$ as can be seen from the presence
of $y_\perp^2$ in the denominators of the $g_{ij}$-terms.
This indicates that neither $F_-$ nor $F_+$ by itself is sufficient to describe the
geometry of a global Finsler space.
In fact, the first axiom of Finsler functions requires infinite differentiability away from 
$y=0$, a condition that is clearly violated by both $F_+$ and $F_-$.

To begin an analysis of the singular behavior of $F_\pm,$ recall that in conventional Finsler geometry the Finsler function $F$ scales as $F (\la y) = \la F(y).$  The scaling property and reparametrization of the  distance functional
\beq
D_F = \int F(y(\la)) d \la,
\eeq
imply that any path can be reparamterized to lie on the level set defined by $F=1.$  The corresponding hypersurface is called the {\it indicatrix:} the indicatrix suffices to investigate the geometry of the Finsler space.

To proceed, we plot both $F_+$ and $F_-$ and construct an associated indicatriix for each, the resulting plot exhibits 
smooth paths from the hypersurface associated to $F_-$ to  the hypersurface associated to $F_+$
at the points where the geometry of either one becomes singular.  
This suggests that the $F_+$ and $F_-$ should be considered as arising from a single
algebraic variety that might be desingularized prior to root extraction.
It is then consistent to impose the condition $F=1$ with the result that the variety is now
a double-cover of the sphere.  This is checked explicitly below.

Squaring twice to eliminate the square roots in Eq.(\ref{origfins}) yields the polynomial condition $f(F,y_p,y_\perp^i)=0$ that defines an algebraic
variety $X \subset \mathbb{R}^{n+1}$, with
\beq
f(F,y_p,y_\perp^i) = (F^2-y^2)^2 - b^2 y_{\perp}^2(2 (F^2+ y^2)-b^2 y_\perp^2)  = 0.
\label{origvariety}
\eeq
The variety is invariant under the transformation $(F \rightarrow \la F, y^i \rightarrow \la y^i)$,
the generalization of the Finsler function homogeneity condition. 
Note that the gradient of $f$ vanishes when $y_\perp^i = 0$, indicating the presence of a
singular set $\Si$, a line in the variety $X$. In particular, $X$ is not a smooth manifold as it stands.
Note that the above procedure introduces $F < 0$ solutions, however, these 
do not intersect the $F>0$ variety and can therefore be treated independently.
The indicatrix can now be constructed by setting $F=1$, now possible since $F$ is expressed in terms of the variety $X$.
This results in the constraint
\beq
y^2 = \left(1 \mp  \sqrt{b^2 y_\perp^2} \right)^2,
\label{pertcirc}
\eeq
on the $y^i$ variables,
and the two solutions correspond to the two roots $F_\pm$.
Eq.(\ref{pertcirc}) explicitly exhibits the indicatrix as a double-valued, small perturbation from the 
half-circle in the $(|y_\perp|,y_p)$ half-plane.
The singular points result at the poles $y_\perp=0$ where the derivatives 
$y_p^\prime(|y_\perp|) \rightarrow \pm 2 |b|$ 
fail to vanish, a condition required for the hypersurface of revolution (symmetric
in the $y_\perp^i$) to be smooth.

\vglue 0.6cm
{\bf \noindent III. FORMAL DESINGULARIZATION}
\vglue 0.4cm

For simplicity in notation, only the singular points where $F>0$ and $y_p>0$ are considered
in what follows, as the others can be handled similarly using the symmetries of the defining 
variety.
To desingularize the variety, define a new coordinate $u^i$ so that
\beq
y_\perp^i = (F^2 - y_p^2) u^i
\label{ucoords}
\eeq
 and study the new variety in a small neighborhood of 
the singular point.
Plugging Eq.(\ref{ucoords}) into the variety equation yields
$f = (F^2 - y_p^2)^2 h(F,y_p,u^i)$,
where 
\beq	
h = (1 - (F^2 - y_p^2) u^2)^2 - b^2 u^2 \left[ 2 (F^2 + y_p^2) 
+ (2 - b^2) (F^2 - y_p^2)^2 u^2 \right] ,
\eeq
and $h(F,y_p,u^i) = 0$ yields the same variety away from the singular set.
The exceptional locus $F^2 = y_p^2$ intersects the variety $h=0$ on a sphere (for fixed $y_p$) with
\beq
u^2 = {1 \over 4 b^2 y_p^2},
\eeq
demonstrating that the new $u^i$ variables are not all identically zero at the singular points.
A new desingularized variety $\tilde X$ can then be defined by first removing all points in a
small neighborhood of the singular set
from $X$ and then "gluing in" a copy of $S^{n-2} \times \mathbb{R}$,
where the line $\mathbb{R}$ is just $y_p$ and $S^{n-2}$ is the sphere determined by $u^i$
at the appropriate fixed $y_p$ value.
Then $\tilde X$ admits a smooth, differentiable structure.

To identify the neighborhood about the singular point for which the $u^i$ coordinates are valid,
general solutions to $F^2 - y_p^2 = 0$ can be examined.
In addition to $y_\perp^i = 0$, there is another solution on the $F_-$ sheet given by
\beq
y_\perp^2 = {4 b^2 \over (1 - b^2)^2} y_p^2  \equiv \ep^2.
\label{secondsol}
\eeq
The neighborhood in which the $u^i$ coordinates give a smooth structure is therefore restricted to the region $y_\perp^2 < \ep^2$ so that 
the transformation (\ref{ucoords}) is non-singular.
This also happens to be precisely the region where $F_-$ fails to be convex.
Note that the gradient of $h$ is nonzero everywhere in this region with the singular set 
lifting to a sphere in the new coordinates. 
This is commensurate with what is expected from the spin variable
in $n=4$ dimensions where the singular set corresponds to a two-sphere on which the
classical spin of the particle points.

The variables $u^i$ can be thought of as auxiliary variables that specify a smooth gluing
of the two sheets present after the singular set is removed.
Expressing $u^i$ in terms of $y^i$ gives
\beq
u^i = {y_\perp^i \over (F^2 - y_p^2)} = 
{ y_\perp^i \over (1 + b^2) y_\perp^2 \pm 2 |b| |y_\perp|\sqrt{y^2} },
\eeq
yielding two solutions, one for the outer sheet determined by $F_+$ and the other
for the inner sheet determined by $F_-$.
At the singular point $u^i$ is the vector that points along $y_\perp^i$ for the positive
choice and opposite to $y_\perp^i$ for the negative choice.
A continuous curve $\gamma(t)$ in the variety $\tilde X$ must therefore change 
sheets in $X$ as it passes through a singular point so that $u^i(t)$ remains continuous as $y_\perp^i$ necessarily changes sign.

\vglue 0.6cm
{\bf \noindent IV. IMPROVED COORDINATES AND A NEW METRIC}
\vglue 0.4cm

Near the singular point, the metric calculated using the $y^i$ variables diverges as
is seen in Eq.\ (\ref{metric1}).  It is natural to ask if the $u^i$ variables can
be used to define a finite, consistent metric in the neighborhood of the singular point. 
Solving for $F$ in terms of the $u$-variables gives
\beq
\tilde F_\pm(y_p,u^i) = {1 \over (1 - b^2)\sqrt{u^2}} \left[ \sqrt{1 + (1-b^2)^2 u^2 y_p^2} \pm |b| \right] .
\eeq
Only the lower sign has the correct limit at the singular point, so it can be deduced that
$F_u = \tilde F_-$ is the relevant Finsler function.
Note that the Finsler function as expressed in terms of the new variables is in fact smooth and
single-valued near the singular point as expected from the desingularization.
Unfortunately, $F_u(y_p,u^i)$ is not a homogeneous function of its new variable set as 
$u^i \rightarrow (1/ \la) u^i$ when $y^i \rightarrow \la y^i$, so it is not possible to 
use the conventional argument to define a Finsler metric.
Fortunately, this problem can be remedied by defining a new variable 
\beq 
w^i = {u^i \over u^2} = {F^2 - y_p^2 \over y_\perp^2} y_\perp^i,
\eeq
with the same scaling properties as $y_i$.
Then $F_u$ becomes a homogeneous function in terms of $w$ 
\beq
F_u = \sqrt{y_p^2 + {w^2 \over (1 - b^2)^2}} -  \sqrt{b^2 {w^2 \over  (1 - b^2)^2}}.
\eeq
One further obvious scaling $w^i = (1 - b^2)z^i$ brings the Finsler function into the same form
as $F_-$ given by Eq.(\ref{origfins}), one of the original functions we started with.  
In fact, the transformation of variables is a symmetry of the original polynomial condition
$f(F, y_p, y_\perp^i) = 0$ as can be easily verified by direct substitution of $u(z)$ into 
$h(F, y_p, u^i)$
\beq
h(F, y_p, u^i(z)) = {1 \over (1 - b^2)^2 z^4} 
\left[ (F^2 - y_p^2 - z^2)^2 - b^2 z^2(2(F^2 + y_p^2 + z^2) - b^2 z^2)\right] ,
\eeq
where the factor in brackets is just $f(F,y_p,z^i)$, the function defined in Eq.(\ref{origvariety})
with $y_\perp^i \rightarrow z^i$.

A relevant fact is that the original singular point $y_\perp^i = 0$ 
maps to a sphere of radius $\epsilon  = 2 |b| y_p /(1 - b^2)$ in the $z^i$ variables, while
the second solution to $F_-^2 = y_p^2$ given in Eq.(\ref{secondsol}) for
$y_\perp$ gets mapped to a singular set in the new coordinates, $\Si^\prime$, defined
by $z^i = 0$.
The metric can now be computed exactly as it is for $F_-$ in terms of $z$ and gives a finite, 
well-defined value everywhere on this sphere.
Writing $z^i$ in terms of the original variables using $F_\pm$ yields
\beq
z^i = {1 \over 1 - b^2} \left({F^2 - y_p^2 \over y_\perp^2} \right) y_\perp^i= {1 \over 1 - b^2} \left[ 1 + b^2 \pm 2 {\sqrt{ b^2 y^2 \over y_\perp^2}}\right] y_\perp^i ,
\eeq
demonstrating explicitly that $z^i$ survives at the singular point $y_\perp^i \rightarrow 0$ with 
\beq
z^2 \rightarrow {4 b^2 y_p^2 \over (1 - b^2)^2},
\eeq
and direction parallel to the unit vector along $y_\perp^i$ as the limit is taken.
Calculation of the metric using $F_-(y_p,z^i)$ on the singular set yields
\beq
\tilde g_{pp} = 1 - {8 b^4 \over (1 + b^2)^3} , \quad 
\tilde g_{pi} = - |b| {(1 - b^2)^3 \over (1 + b^2)^3} \hat z^i , \quad
\tilde g_{ij} = {(1 - b^2)^2 \over 2 (1 + b^2)} \left[ \delta_{ij} 
+ {(1 - b^2)^2 \over (1 + b^2)^2} \hat z^i \hat z^j \right] .
\eeq
Note that the result is finite and well-defined and depends only on the unit vector $\hat z$ on the sphere.

Together, the two charts on $\tilde X$ defined using $\{F, y_p, y_\perp^i\}$ on $X - \Si$ 
and $\{F , y_p, z^i\}$ on $X - \Si^\prime$ form an atlas for the de-singularized variety $\tilde X$.  
Note that it is always possible to chose one or the other set of coordinates to enforce the 
condition $\det{g} \ge 0$.
The only issue is that there exists a sphere in the space 
(on $F_-$ at $y_\perp = \sqrt{1 - b^2} ~ \ep/2$ in terms of the $y^i$ variables) where $\det{g}$ is identically zero
for either choice of charts.
This indicates that the resulting global Finsler structure is only positive semi-definite, not strictly
positive definite.  
The physical meaning of this result is discussed in the next section.

\vglue 0.6cm
{\bf \noindent V. LAGRANGIAN}
\vglue 0.4cm

The Euclidean structure can be converted back to the original Minkowski structures from
which they were derived
by performing a Wick rotation where $n=4$ dimensions are used and the time-components of the 
four-vectors are multiplied by $i$.
The squares and dot-products of the four-vectors convert over to their Minkowski counterparts
with a sign.  For example, $y^2 \rightarrow - u^2$ where $u^2 = (u_0)^2 - \vec u^2$ 
depends on the Minkowski metric.  
Under this map, $F \rightarrow -i L$, and $y_p \rightarrow u_p$, $y_\perp \rightarrow u_\perp$
where the parallel and perpendicular components of $u$ are determined by the Minkowski
metric.  
Explicitly, $u^\mu = u^\mu_p + u^\mu_\perp$ with (note that this map only works when
$b^2 \ne 0$)
\beq
u_\perp^\mu = u^\mu - {u \cdot b \over b^2} b^\mu .
\eeq
The lagrangian per unit mass (with $b^\mu$ in units of the mass $m$) becomes
\beq
L_\pm = - \sqrt{u^2} \mp \sqrt{b^2 (- u_\perp^2)} .
\eeq
The desingularization variable $z^i$ maps to 
\beq
z^\mu = {1 \over 1 + b^2} \left[ {L^2 - u_p^2 \over u_\perp^2} \right] u_\perp^\mu = 
{1 \over 1 + b^2} \left[ 1 - b^2  \pm 2 \sqrt{b^2 u^2 \over - u_\perp^2} \right] u_\perp^\mu .
\eeq

The condition that there is a singular point on the variety is now that $u_\perp^\mu = 0$.
When $b^\mu$ is spaclike $(b^2 < 0)$,  the singular point lies in a space-like velocity region that is inaccessible to physical particles making the singular points largely irrelevant.

When $b^\mu$ is time-like $(b^2 >0)$, a Lorentz transformation can be used to go to a frame in which only $b^0 \ne 0$.  In this frame, the singular point is where the three-velocity of 
the particle vanishes, $\vec u = \vec v = 0$.
Attention is therefore focused on the special case $b^\mu = (b^0, 0,0,0)$ so that the
singular point lies at $\vec v = 0$ in what follows.  In addition, standard proper time parametrization
is assumed so that $u^2 = 1$ can be imposed.
In this case, the sign appearing in the definition of $L_\pm$ indicates the velocity-helicity of the
particle, and it must be prescribed in order to define a particle's trajectory.
Note that velocity-helicity and momentum-helicity can be different due to the
modification of the momentum-velocity relation
\beq
p^i = \ga v^i \mp b_0 \hat v 
\eeq

It is not surprising that the metric breaks down at $\vec v = 0$ since the three-momentum tends to
$\vec p \rightarrow \mp b_0 \hat v \ne 0$, 
so the direction of the momentum is not determined there.
The desingularization variable at the singular point is
\beq
z^\mu \rightarrow \left( 0, \pm {2 b_0 \over 1 + b_0^2} \hat v \right ) ,
\eeq
where $\hat v$ is a unit vector that can point in any spatial direction.
The positive and negative values for $z$ can be associated with the velocity-helicity of
the particle, therefore the sign is determined by the particle's spin direction, $\pm \hat v$.
If $\vec z$ is required to vary continuously as a particle moves through the singular point
(by stopping and reversing direction)
then the momentum must also remain continuous along the trajectory indicating that the
transition from $L_\pm$ to $L_\mp$ is required.  

The region where the Finsler metric of the 
Euclidean case fails to be positive definite corresponds to the low-speed
region where $\vec v^2 \le b_0^2 / (1 + b_0^2)$ and the momentum is in fact opposite the velocity.
In this region it is possible to increase the particle's momentum while decreasing its velocity and 
energy ($p^0 = \ga m$ can be expressed purely in terms of the velocity).

Another curious implication of the vanishing of certain eigenvalues of the metric in directions 
orthogonal to the particle velocity is the existence of extremal action solutions with $\vec p = 0$
that have nonzero velocity.  Arbitrary changes in the direction of the velocity do not change the 
action indicating that the direction of $\vec v$ can randomly vary as the particle moves along.
The particle can even exhibit uniform circular motion with zero force.
These types of motion should be considered spurious and indicate a problem with the model, 
presumably due to the fact that spin has been fixed to helicity eigenstates in this limit of the
full quantum theory.  Desingularization can smooth out the variety at the singular point and 
indicate the source of these spurious trajectories.  At the singular point it is not sufficient to 
specify the velocity but necessary to also retain some unit vector direction proportional to 
the momentum vector.  This suggests there is an additional internal variable in the system
determining the direction of this momentum, namely the particle spin. Requiring that the
particle spin angular momentum be conserved is one way to eliminate these
spurious trajectories.  

\vglue 0.6cm
{\bf \noindent VI. GENERAL PROCEDURE}
\vglue 0.4cm
In this section, we make several remarks concerning a general procedure 
that could be used to perform the desingularization within
the context of the general derivation presented in \cite{kr}.
The procedure outlined is \cite{kr} starts with a general SME fermion dispersion relation that 
yields a generic polynomial condition $P(L) = 0$,
where the coefficients in the polynomial depend on $u^\mu$, $m$, and the Lorentz-violating 
parameters. 
The solutions of this equation determine the possible lagrangians as perturbations
of the conventional $L = \pm m \sqrt{u^2}$ structure.
The four degrees of freedom of the relativistic theory split into two particle and two anti-particle
states according to the overall sign of this Lagrangian.
This fact can be deduced by examination of the defining equation $L = - p_\mu u^\mu$
in the rest frame where $\vec u = 0$, recalling that  the negative-energy solutions are 
reinterpreted at the level of the quantum field theory as antiparticles.
The desingularization procedure is applied to the corresponding variety $f(F,y^i)=0$
obtained from $P(L)=0$ by using an appropriate Euclideanization procedure, like Wick rotation.
The singular points are then determined by finding points where the gradient of $f$ is zero.
An appropriate blowup procedure should then be used to produce a smooth manifold
that sits above the variety.  
In this paper, the blowup is performed using an additional set of coordinates near the
singular point that are related to the slopes of lines through the singular point. 
Using appropriate new coordinates in the region where the original metric
fails to be positive definite, it is possible to define a new metric that is positive semi-definite
everywhere on the manifold.
In general, finding an appropriate set of "new coordinates" near the singular points
should be possible.
The existence of suitable coordinates that make the resulting quadratic form semi-positive
definite is an interesting open issue.

In addition to the physical solutions, there are often additional, spurious solutions to 
$P(L)=0$ that yield nonperturbative values for the lagrangian that do not have the
correct standard limit as the Lorentz-violation parameters are tuned to zero.
For example, in the $b^\mu$-case, $P(L)$ is an octic polynomial and there is an additional 
solution $L = - \alpha \sqrt{(b \cdot u)^2}$, with $\alpha = \sqrt{m^2  / b^2 + 1}$ that leads to a zero metric in the corresponding Finsler space.
The indicatrix is the flat hypersurface $y_p = 1/\sqrt{1 - b^2}$ that is tangent to the top of the $F_- = 1$ singular Finsler function.
In fact, this solution can be applied in the region where $F_-$ fails to be convex,
rather like the Gibbs construction of thermodynamics. 
Physically, the spurious lagrangian corresponds to particles that have 
zero velocity and the fixed momentum value $p_\mu = \al \sign(b \cdot u) b_\mu$.
Applying this "Gibbs construction" to time-like $b \rightarrow (b^0,0,0,0)$ case 
in the region where $F_-$ fails to be convex yields a region at low velocities 
where the particle with negative velocity-helicity gets mapped to a zero velocity state.
In this case, the badly-behaved region is simply collapsed to a point.
Despite this curious mathematical possibility, it seems unlikely to correspond to a
realistic model as the spin would somehow have to constantly adjust to keep the velocity zero at these very low velocities violating spin angular momentum conservation.

\vglue 0.6cm
{\bf \noindent VI. CONCLUSIONS}
\vglue 0.4cm

Desingularization of the lagrangian variety that arises from the SME $b^\mu$-coupling 
can be accomplished by first converting to Euclidean space where the theory is determined
by a pair of singular Finsler functions.
These functions are defined using an algebraic variety which is then desingularized 
using a new parametrization of the variety.
This procedure yields an implicitly determined $F$ on an underlying manifold.
The fact that $F$ is determined implicitly by a variety condition and is generally double-valued
violates the first axiom of Finsler geometry which states that $F$ must be a function.
In addition, there are regions where the metric becomes negative definite near the singular
point.
It turns out to be possible to parameterize the "badly behaved" 
It would be natural to generalize the concept of a Finsler geometry to allow an appropriate
formulation on the lifted variety (a manifold), but for 
the single remaining impediment due to the degeneracy of the metric
at very specific values of the velocity.
This degeneracy corresponds to a spurious set of extremal paths at low velocity.
In the physical case, these paths might be inaccessible to the particle (as is the case for space-like
$b$) or may be eliminated by requiring conservation of spin, a
variable that has been discarded in the transition to the classical model.
The combination of the two charts forms an atlas on the resulting manifold that allows 
identification of smooth paths through the singular set in the original variety.
The result is a smooth classical model on a manifold that sits above the variety that can be
used for practical calculations.  

The explicit construction for the $b^\mu$ case presented in this paper suggests that it is 
possible to desingularize classical lagrangians defined on singular algebraic varieties through standard procedures yielding consistent models for the classical ray limit of Lorentz-violating theories.
Such a model would yield a consistent formulation of classical particle propagation within the context of general relativity when explicit Lorentz violation is present in the matter sector.
Some generalization of pseudo-Riemann-Finsler geometry which involves a formulation on a 
variety and allows for degenerate metric directions inside the light cone appears to be required
to consistently describe these models in a geometric way.
These new requirements are a fundamental implication of including spin coupling into the theory
as a physically relevant quantity.

\vglue 0.6cm
{\bf \noindent ACKNOWLEDGMENTS}
We wish to acknowledge the support of New College of Florida's faculty
development funds that contributed to the successful completion of
this project.

\vglue 0.6cm
{\bf\noindent REFERENCES}
\vglue 0.4cm

\end{document}